\documentclass[letterpaper]{article}

\usepackage[T1]{fontenc}

\usepackage{geometry}
\usepackage{setspace}

\usepackage[
  style=chem-acs,
  articletitle=true
]{biblatex}
\usepackage{graphicx}
\usepackage{float}
\newfloat{scheme}{htbp}{los}
\floatname{scheme}{Scheme}
\floatname{chart}{Chart}
\newfloat{graph}{htbp}{loh}

\usepackage{chemformula} 
\usepackage[version = 4]{mhchem} 

\usepackage{authblk}
\author[1,$\dagger$]{Sharad Kumar Yadav}
\author[2,$\dagger$]{Luca Nessi}
\author[3]{Ondrej Dyck}
\author[4]{Jinchen Wang}
\author[3]{Bogdan Dryzhakov}
\author[3]{Alex Melendez}
\author[3]{Huan Zhao}
\author[2]{Qian Song}
\author[1] {Doha Amer}
\author[4]{Cole Brabec}
\author[1]{Saleh Alqazlan}
\author[4]{Ruonan Han}
\author[2]{Riccardo Comin}
\author[3]{Stephen Jesse}
\author[4]{Dirk Englund}
\author[1,4,*]{Jawaher Almutlaq}

\affil[1]{Materials Science and Engineering and Applied Physics, King Abdullah University of Science and Technology, Thuwal, 23955, Saudi Arabia}

\affil[2]{Department of Physics, Massachusetts Institute of Technology, 50
Vassar St, Cambridge, 02139, MA, USA}

\affil[3]{Center for Nanophase Materials Sciences, Oak Ridge National Laboratory, Oak Ridge, TN, USA}

\affil[4]{Research Laboratory of Electronics, Massachusetts Institute of Technology, 50 Vassar St, Cambridge, MA, 02139, United States}

\title{Foundry CMOS platform for multimodal quantum materials characterization}
\date{*Email: Jawaher.Almutlaq@kaust.edu.sa}

\affil[$\dagger$]{These authors contributed equally to this work.}

\begin{document}

\maketitle

\begin{abstract}
Quantum materials experiments increasingly rely on microwave, electrical, thermal, optical, and structural probes, but these capabilities are typically assembled from custom hardware that limits reproducibility and scalability. Here we show that a commercial 65-nm CMOS process can be repurposed as a passive, foundry-manufacturable characterization platform by functionally partitioning its metal stack into microwave, thermal, and electrical subsystems within a 1~mm\textsuperscript{2} footprint. The integrated RF architecture enables cryogenic magnetic susceptibility measurements of Fe$_3$GeTe$_2$ heterostructures at 1.75 K without sample-specific fabrication. We further demonstrate NV-center optically detected magnetic resonance (ODMR) with >20\% contrast at 4--9 dBm microwave power, reducing power requirements by 20-25 dB relative to conventional antenna-based approaches while maintaining sensitivities of 2-3 $\mu$T/$\sqrt{\mathrm{Hz}}$. We additionally confirm compatibility with in-situ electron-beam imaging, showing no measurable degradation in image quality upon device operation. These results establish a scalable, foundry-manufacturable platform for multimodal quantum sensing and materials characterization.

\end{abstract}

\\

Modern quantum materials experiments increasingly rely on microwave, electrical, thermal, optical, and structural probes. These capabilities are typically assembled from discrete hardware components, requiring experiment-specific integration that limits reproducibility, increases system complexity, and complicates measurements of air-sensitive materials under cryogenic conditions \cite{deiseroth2006fe3gete2,powalla2023seeding}. As multimodal characterization becomes increasingly important for studying emergent quantum phenomena, scalable platforms that support multiple measurement modalities within a unified architecture are needed. Recent perspectives have further emphasized that progress in quantum
materials research increasingly depends on correlating device performance
with complementary materials characterization techniques and developing
reproducible experimental workflows~\cite{deLeon2021}. This need is reinforced by recent efforts toward autonomous and AI-assisted materials research, which depend on standardized, reproducible experimental hardware capable of integrating complementary measurement modalities \cite{Stach2021Matter, Wu2022Operando}. Foundry manufacturing enables reproducible hardware platforms that can be deployed across laboratories and applications. Extending this paradigm to quantum materials characterization could reduce experiment-specific hardware development while improving reproducibility and scalability.

Recent advances have addressed individual aspects of this challenge \cite{10.1063/5.0160321}. NV-center quantum sensors provide nanoscale magnetic imaging \cite{rondin2014magnetometry, Schirhagl2014, ,Maze2008}, cryogenic CMOS systems enable compact microwave control and readout \cite{ibrahim2020high, kim2019cmos, fakkel2024cryo}, and in-situ microscopy techniques offer simultaneous structural and functional characterization \cite{BarahDavidJoseph2025, 10.31399/asm.cp.istfa2025p0567, Li2026}. However, these capabilities remain primarily confined to specialized platforms, and their integration within a foundry-manufacturable architecture remains largely unexplored. From these recurring experimental needs, we identify five system-level requirements for scalable quantum materials characterization: (i) efficient and programmable field generation, (ii) electrical and thermal interfacing, (iii) optical accessibility, (iv) compatibility with structural imaging techniques, and (v) reproducible fabrication enabled by commercial semiconductor manufacturing. Together, these requirements define a common hardware framework that spans quantum sensing, cryogenic materials characterization, and in-situ microscopy workflows.

Here, we present a foundry-fabricated CMOS platform designed to satisfy these requirements within a unified architecture. Implemented in a commercial 65-nm RF CMOS process, the platform integrates microwave delivery, electrical interfacing, and thermal control within a compact 1~mm$^2$ footprint while remaining compatible with cryogenic operation down to 1.75 ~K, optical spectroscopy, and electron microscopy. We validate the platform through three representative applications: low-power NV-center quantum sensing, cryogenic magnetic susceptibility measurements of Fe$_3$GeTe$_2$ heterostructures without sample-specific lithography or device fabrication, and direct operation under electron-beam imaging conditions. The integrated RF architecture enables ODMR contrast of 20--26\% at microwave powers of only 4--9~dBm, corresponding to a 20--25~dB reduction in required power relative to conventional antenna-based approaches while maintaining comparable sensitivity. Together, these results establish a scalable and reproducible CMOS hardware platform for multimodal quantum sensing and quantum materials characterization.

\begin{figure*}[!ht]
    \centering
    \includegraphics[width=1\textwidth]{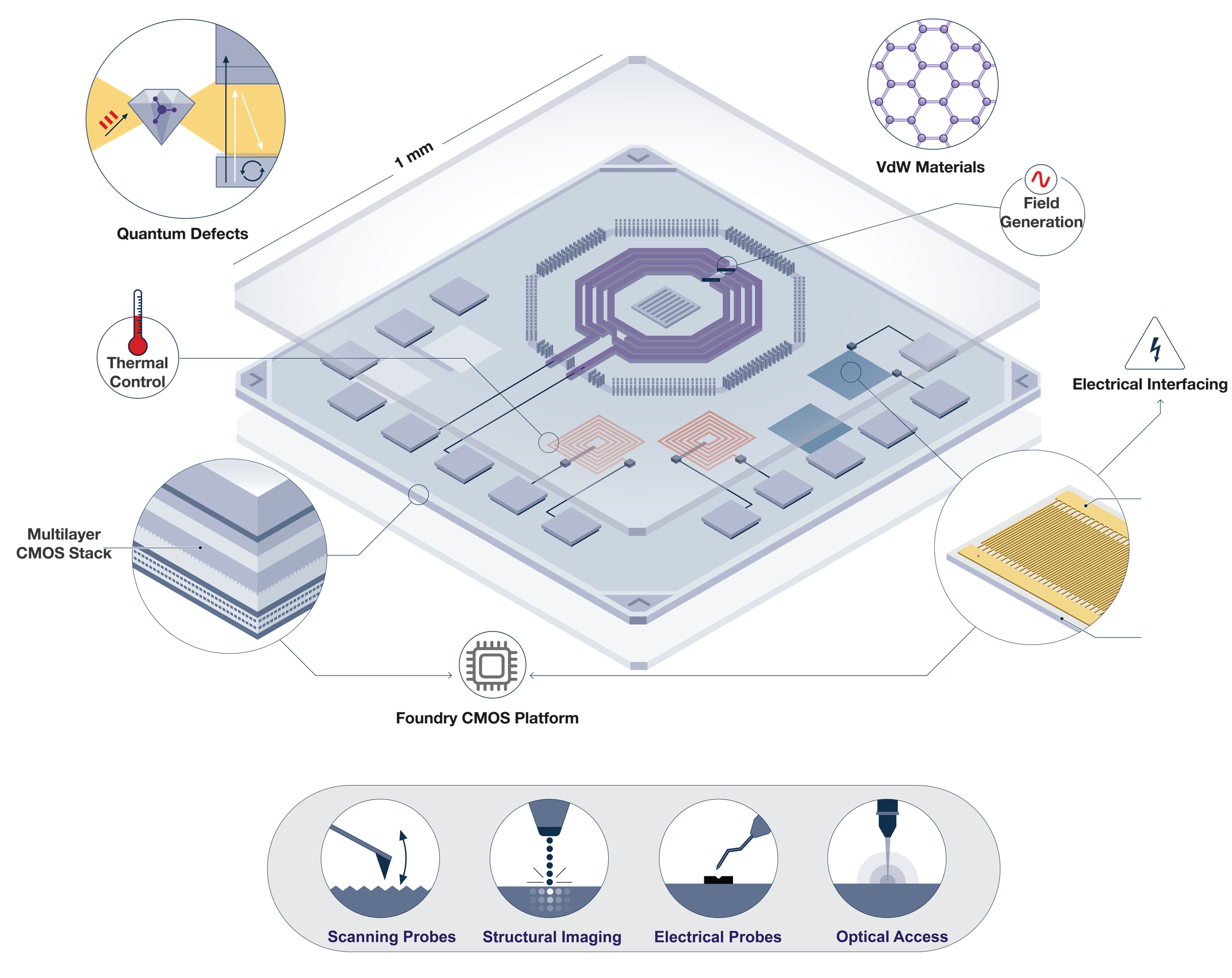} 
\caption{\textbf{Foundry-fabricated CMOS platform for multimodal materials characterization.}
Schematic illustration of the CMOS-integrated platform implemented in a commercial semiconductor process. The architecture supports the heterogeneous integration and multimodal characterization of quantum materials. Insets highlight key platform components, including the integrated RF inductor, resistive heater, interdigitated electrodes (IDE), layered quantum materials integrated onto the chip surface, and the underlying CMOS stack. The bottom panel summarizes representative characterization modalities enabled by the platform, including scanning-probe measurements, structural imaging, electrical probing, and optical spectroscopy.}

\end{figure*}
    \label{fig:CMOS_platform}
    
\section*{Concept and platform architecture}

The platform is implemented in a commercial TSMC 65-nm RF CMOS process, leveraging a technology originally developed for highly integrated mixed-signal and radio-frequency systems. A key advantage of this node is its heterogeneous back-end-of-line stack, which combines up to nine metallization layers, ultra-thick top-metal options, Metal-insulator-metal/metal-oxide-metal (MiM/MoM) passive components, and underlying active devices within a standardized foundry process. Such features have enabled the realization of high-performance passive structures including inductors, transmission lines, capacitors, antennas, and millimeter-wave circuits in prior work, highlighting the versatility of the technology as a substrate for electromagnetic functionality rather than solely digital electronics \cite{lourandakis2014integrated,farahabadi2013compact}. In this work, we exploit these process capabilities to implement a multimodal materials characterization platform through functional partitioning of the CMOS stack.

\begin{figure*}[!htp]
    \centering
    \includegraphics[width=1\textwidth]{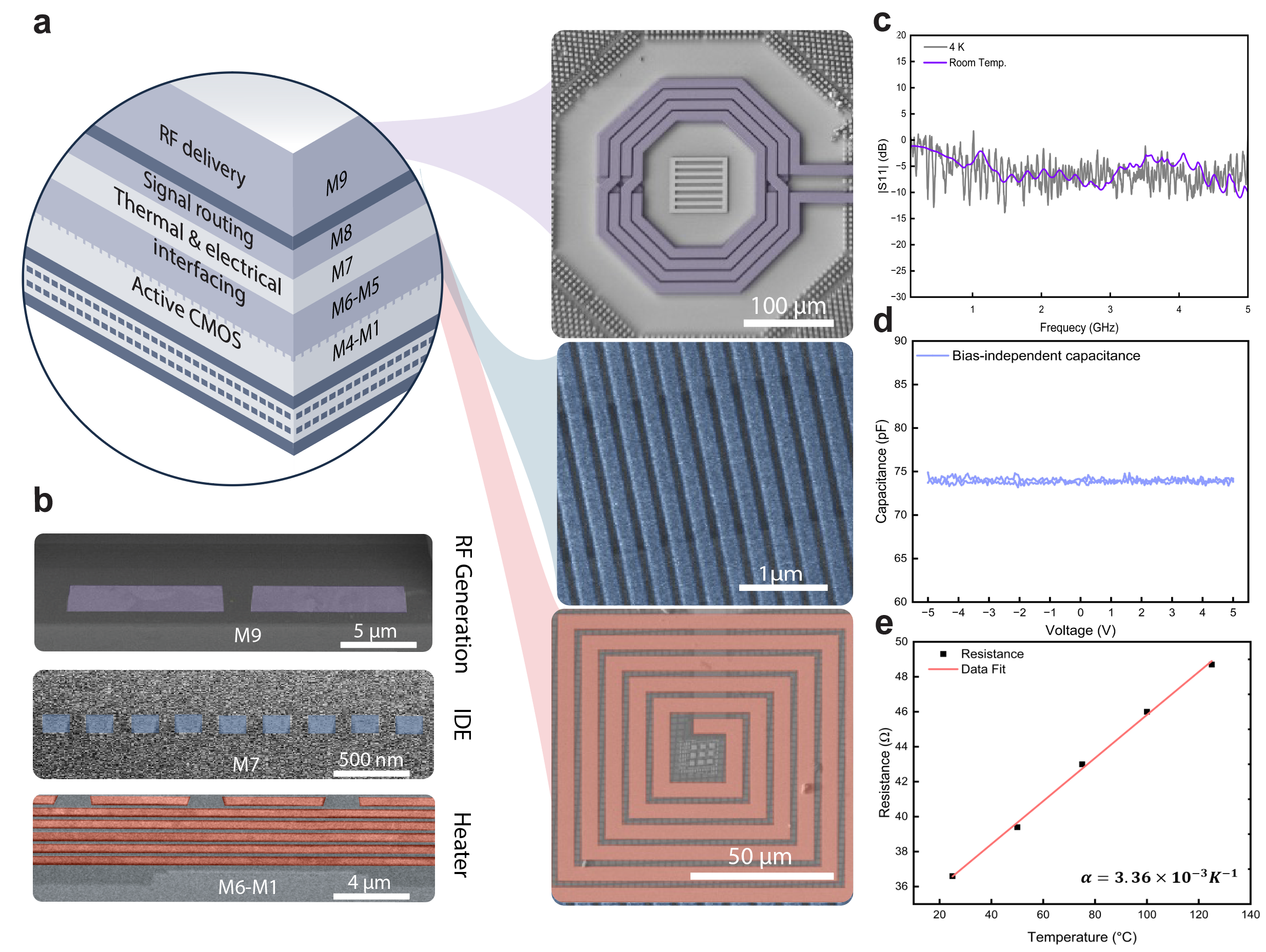} 
\caption{\textbf{Fabricated CMOS-integrated platform: architecture and subsystem characterization.}
(a) Schematic of the layered chip stack and corresponding functional partitioning of the CMOS metal layers, together with false-colored SEM images of the realized structures: the octagonal RF inductor (100~$\mu$m scale bar), the interdigitated-electrode (IDE) array (1~$\mu$m scale bar), and the resistive heater (50~$\mu$m scale bar). (b) False-colored cross-sectional SEM images of the constituent components, showing the RF structure, IDE, and heater within the CMOS stack. (c) Measured S-parameters of the RF structure, showing the reflection coefficient ($S_{11}$) across the 1--5~GHz band. (d) Measured capacitance--voltage (C--V) response of the IDE during forward and reverse voltage sweeps from $-5$ to $+5$~V, showing a bias-independent capacitance of $73.9$~pF with a standard deviation of $0.3$~pF over the measured voltage range. (e) Extracted heater resistance as a function of temperature, yielding a temperature coefficient of resistance $\alpha = 3.36 \times 10^{-3}~\mathrm{K}^{-1}$.
}
\end{figure*}

\subsection*{RF architecture and performance}

Microwave field delivery is implemented using a four-turn octagonal on-chip loop inductor with an outer span of $212~\mu$m, an inner opening of $120~\mu$m, a conductor width of $10~\mu$m, and an inter-turn spacing of $2~\mu$m. Electromagnetic simulations indicate that the geometry concentrates the RF magnetic field within the central active region, enabling localized magnetic excitation (\textbf{Fig.~2a)}. The RF response was characterized through broadband S-parameter measurements from $10~\mathrm{MHz}$ to $5~\mathrm{GHz}$, followed by matrix-based de-embedding to isolate the intrinsic device behavior. The de-embedded response exhibits a return loss ($S_{11}$) below $-5~\mathrm{dB}$ across most of the measured bandwidth, corresponding to more than $68\%$ coupling of incident power into the device. The insertion loss ($S_{21}$) remains close to -1 dB at 4 K and -3 dB at room temperature across the operating bandwidth, indicating higher power transmission and lower dissipative loss (Supplementary \textbf{Fig.~S1a}). Measurements performed at room temperature and $4~\mathrm{K}$ show only minor variations in the RF response, demonstrating compatibility with cryogenic operation. Together, these results establish efficient and robust RF delivery across the frequency range relevant for spin manipulation and AC magnetic excitation.

\subsection{Thermal control and heater performance}\label{subsec3.3}

Thermal control is implemented using a six-turn meander resistive heater fabricated within the lower metallization layers (M1--M6). The meander geometry provides a compact high-resistance structure that enables efficient Joule heating while maintaining spatial localization of the temperature profile \textbf{(Fig.~2b).} Linear current--voltage characteristics measured between $25~^{\circ}$C and $125~^{\circ}$C confirm ohmic heater operation (Supplementary \textbf{Fig.~S1b}). The thermal response of the heater is characterized through four-probe electrical measurements using the temperature dependence of its resistance, R(T), yielding the temperature coefficient of resistance shown in \textbf{Fig.~2e}. The extracted temperature coefficient of resistance is $\alpha = 3.36 \times 10^{-3}~\mathrm{K}^{-1}$, consistent with metallic conductors \cite{DELLINGER1910213,simon1992properties}. The device-to-ambient thermal conductance is measured to be $G_{\mathrm{th}} = 5.16~\mathrm{mW/K}$, corresponding to a thermal resistance of $R_{\mathrm{th}} \approx 194~\mathrm{K/W}$. These values indicate efficient thermal isolation, enabling controlled local temperature modulation with moderate power dissipation. The integrated heater supports local thermal control over a wide operating range (1.75--393 K), enabling temperature-dependent materials characterization.

\subsection{Electrical interfacing and IDE functionality}

Electrical interfacing is provided through on-chip interdigitated electrodes (IDEs) fabricated in metal layer M7. The IDE consists of alternating metal fingers with widths of 200~nm and inter-finger spacings of 120~nm, occupying an active area of approximately $100~\mu\mathrm{m} \times 100~\mu\mathrm{m}$. This geometry generates localized electric fields near the electrode surface, enabling sensitivity to dielectric perturbations in materials placed in the immediate vicinity of the device. Capacitance-voltage measurements showed a bias-independent baseline capacitance of approximately 73.9~pF over the applied $\pm5$~V range \textbf{Fig.~2d}, consistent with the passive metal--dielectric architecture of the IDE. The electrical response of the IDEs was characterized using complementary voltage-forcing and current-forcing measurements, revealing a low-leakage operating regime with linear current--voltage behavior and stable differential conductance (Supplementary\textbf{ Fig.~S2e,f}). This operating window defines the bias range for reliable electrical interfacing and sensing. Finite-element electrostatic simulations indicate that the electric field is strongly confined near the electrode surface, providing sensitivity to dielectric loading within the near-field region (Supplementary \textbf{ Fig.~S2a,b}). To further enhance this sensitivity, the design utilizes an underlying metal layer as an etch mask to partially suspend the active structures, reducing dielectric loading from the substrate and minimizing parasitic screening effects. The predicted capacitance modulation with varying dielectric permittivity and overlayer thickness (Supplementary \textbf{Fig.~S2c,d}) establishes the expected sensing response of the IDE to changes in the surrounding dielectric environment.

\section{Magnetic susceptibility measurements}

A key challenge in the characterization of low-dimensional magnetic materials is the reliance on sample-specific device fabrication, which introduces variability across experiments and complicates measurements of air-sensitive systems \cite{Fei2018, Lanza2020} To evaluate whether the CMOS platform can address this limitation, we performed Kerr-rotation-based magnetic susceptibility measurements on hBN/Fe$_3$GeTe$_2$/hBN heterostructures transferred directly onto the device without lithographic processing, following the heterogeneous integration approach illustrated in \textbf{Fig. 3a}. An optical micrograph of the completed device is shown in \textbf{Fig. 3b}.

\begin{figure*}[!htp]
    \centering
    \includegraphics[width=1\textwidth]{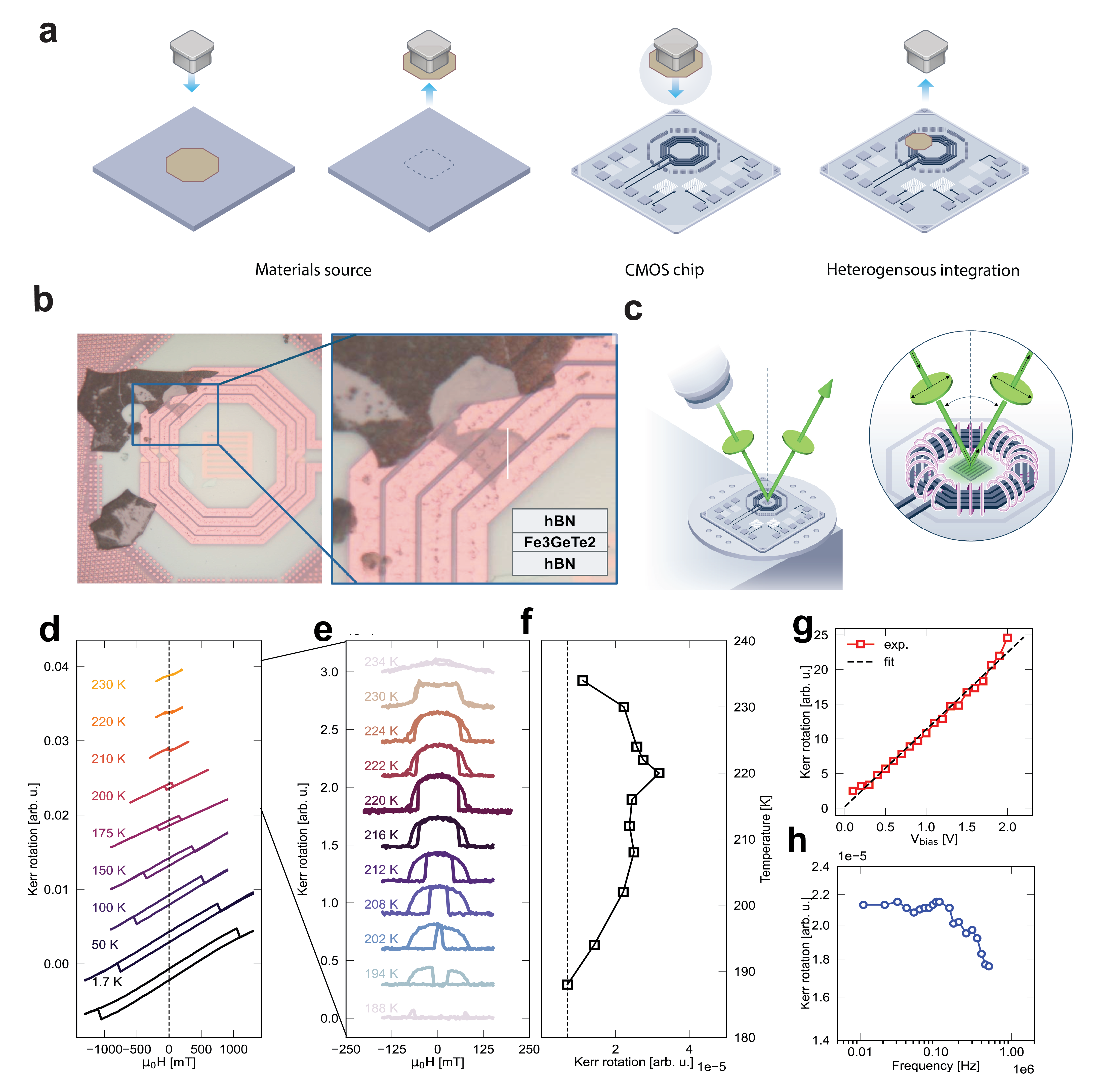}
\caption{\textbf{Cryogenic magnetic susceptibility measurements using the CMOS platform.}
Illustration of the heterogeneous integration process, showing transfer of an encapsulated hBN/Fe$_3$GeTe$_2$/hBN heterostructure from the source substrate onto the foundry-fabricated CMOS characterization chip. (b) Optical micrograph of the CMOS chip after transfer; inset shows the encapsulated heterostructure positioned over the integrated micro-inductor. (c) Schematic of the Kerr-rotation measurement configuration, where the integrated CMOS inductor provides AC magnetic excitation while Kerr rotation is detected optically. (d) Temperature-dependent Kerr-rotation hysteresis loops measured between 1.75 and 230 K. (e) Expanded view of the hysteresis loops near the magnetic phase transition. (f) Temperature dependence of the magnetic susceptibility extracted from the Kerr measurements, showing a maximum near the Curie temperature of Fe$_3$GeTe$_2$. (g) Magnetic susceptibility as a function of inductor bias measured at 220 K, demonstrating a linear response (dashed line: linear fit). (h) Frequency dependence of the susceptibility at 220 K, indicating stable operation across the investigated bandwidth.
}

    \label{fig:magnetic susceptibility}
\end{figure*}

Magnetic susceptibility was measured using Kerr rotation spectroscopy with AC magnetic excitation provided by the integrated micro-inductor \textbf{(Fig.~3c).} The Kerr rotation spectrum exhibits a maximum response near 2.33~eV (533~nm), which was used for subsequent measurements (Supplementary Fig. S3). Representative temperature-dependent Kerr hysteresis loops are shown in \textbf{Fig. 3d}, with an expanded view near the magnetic phase transition in \textbf{Fig. 3e.} An AC magnetic field was generated by driving the on-chip inductor at 111~kHz, while the Kerr response was detected using lock-in techniques synchronized to the excitation frequency. The susceptibility exhibits a pronounced peak near $T \approx 220$~K \textbf{(Fig.~3f)}, consistent with the Curie temperature of Fe$_3$GeTe$_2$ reported in the literature \cite{Fei2018, PhysRevB.93.014411}. Knowing the remanent magnetization value of Fe$_3$GeTe$_2$ from previous literature \cite{may2016magnetic} and the maximum field value generated by the CMOS chip, we conclude that at 220~K we measure a susceptibility of $\chi \simeq \Delta M/\Delta H \simeq 12.2$. Given the signal-to-noise ratio of our measurements, the detection limit for the susceptibility is estimated to be $\sim 0.75$. The measured response scales linearly with the applied inductor bias \textbf{(Fig.~3g),} confirming magnetic excitation by the integrated coil, and remains approximately constant across the investigated frequency range (10--250~kHz), indicating stable operation within the measurement bandwidth \textbf{(Fig.~3h)}. The signal is constant over this range; the roll-off observed near the upper end of the range is attributable to the 3 dB cut-off of the balanced photodetector (bandwidth DC–1 MHz) used for detection, rather than an intrinsic frequency dependence of the sample response.
An additional advantage of the integrated microinductor is its ability to provide dynamic magnetic excitation directly on chip. In the present magnetic susceptibility measurements, the excitation frequency was detector-limited to 250~kHz while remaining stable over the investigated bandwidth. Unlike conventional superconducting magnets, which are primarily optimized for slowly varying or static magnetic fields, the integrated inductor naturally supports high-bandwidth AC excitation and pulsed operation enabled by its RF design. This capability opens opportunities for time-resolved studies, including frequency-dependent magnetic susceptibility, dynamic spin phenomena, and pump--probe measurements, while maintaining localized magnetic field delivery within a compact, foundry-fabricated architecture. Importantly, these measurements are performed without sample-specific lithography or custom-fabricated excitation structures. Encapsulated Fe$_3$GeTe$_2$ flakes are transferred directly onto the passivated CMOS surface under inert conditions, avoiding resist processing and minimizing sample handling.

\section{Quantum control enabled by the CMOS RF architecture}\label{sec5}

ODMR measurements were performed using a commercial scanning NV microscope (Qnami ProteusQ) at room temperature. The measurement configuration is illustrated in \textbf{Fig. 4a}, while \textbf{Fig. 4b} shows an optical micrograph of the CMOS device highlighting the integrated RF inductor used for spin control. The packaged and wire-bonded device used for the experiments is presented in \textbf{Fig. 4c.} To establish a practically relevant benchmark, we compared the integrated CMOS microwave architecture directly against the commercial microwave PCB supplied with the microscope. The measurements were performed on the same instrument using identical optical, microwave, and acquisition settings. The resulting ODMR spectra exhibit resonances in the 2.7--3.0~GHz range.
Electromagnetic simulations \textbf{(Fig. 4d,e)} show that the integrated octagonal inductor generates a localized RF magnetic field that remains appreciable at NV--inductor separations of approximately $10~\mu$m, consistent with efficient near-field microwave delivery. Rabi oscillations measured at 2.764~GHz confirm coherent spin manipulation driven by the integrated RF field  \textbf{(Fig. 4f).} The corresponding ODMR spectra \textbf{(Fig. 4g)} exhibit contrasts of 24--26\% at 9~dBm and remain above 20\% at 4~dBm. In contrast, a commercial microwave antenna requires 24.6--30~dBm to achieve comparable ODMR contrast. This corresponds to a reduction of approximately 20--25~dB (>100$\times$) in transmitted microwave power while maintaining comparable sensing performance. The reduced power requirement arises primarily from near-field localization of the RF magnetic field generated by the on-chip inductor, together with improved power delivery enabled by the integrated RF architecture. The extracted magnetic sensitivity ranges from approximately $2.2$ to $2.9~\mu\mathrm{T}/\sqrt{\mathrm{Hz}}$, with the best performance obtained at 4~dBm owing to reduced linewidth broadening. A direct comparison between the CMOS platform and the commercial antenna is summarized in \textbf{Fig.4h.} The complete power-dependent ODMR measurements are provided in Supplementary \textbf{Fig. S4.} Prior work on CMOS-integrated RF generation for on-chip ODMR has demonstrated the potential of this pathway toward compact, low-power quantum sensing platforms, as summarized in \textbf{Table 1}. Notably, the higher sensitivities reported in some of these platforms stem from optimization on many fronts including optical-detection, larger sensing area and enhanced nanophotonic filtering with differential background rejection rather than from the RF delivery alone \cite{ibrahim2020high, kim2019cmos}. Taken together, this body of prior work illustrates the broader potential of CMOS-integrated architectures to advance quantum sensing and materials characterization, motivating the platform presented here.

\begin{figure*}[htp]
    \centering
    \includegraphics[width=0.8\textwidth]{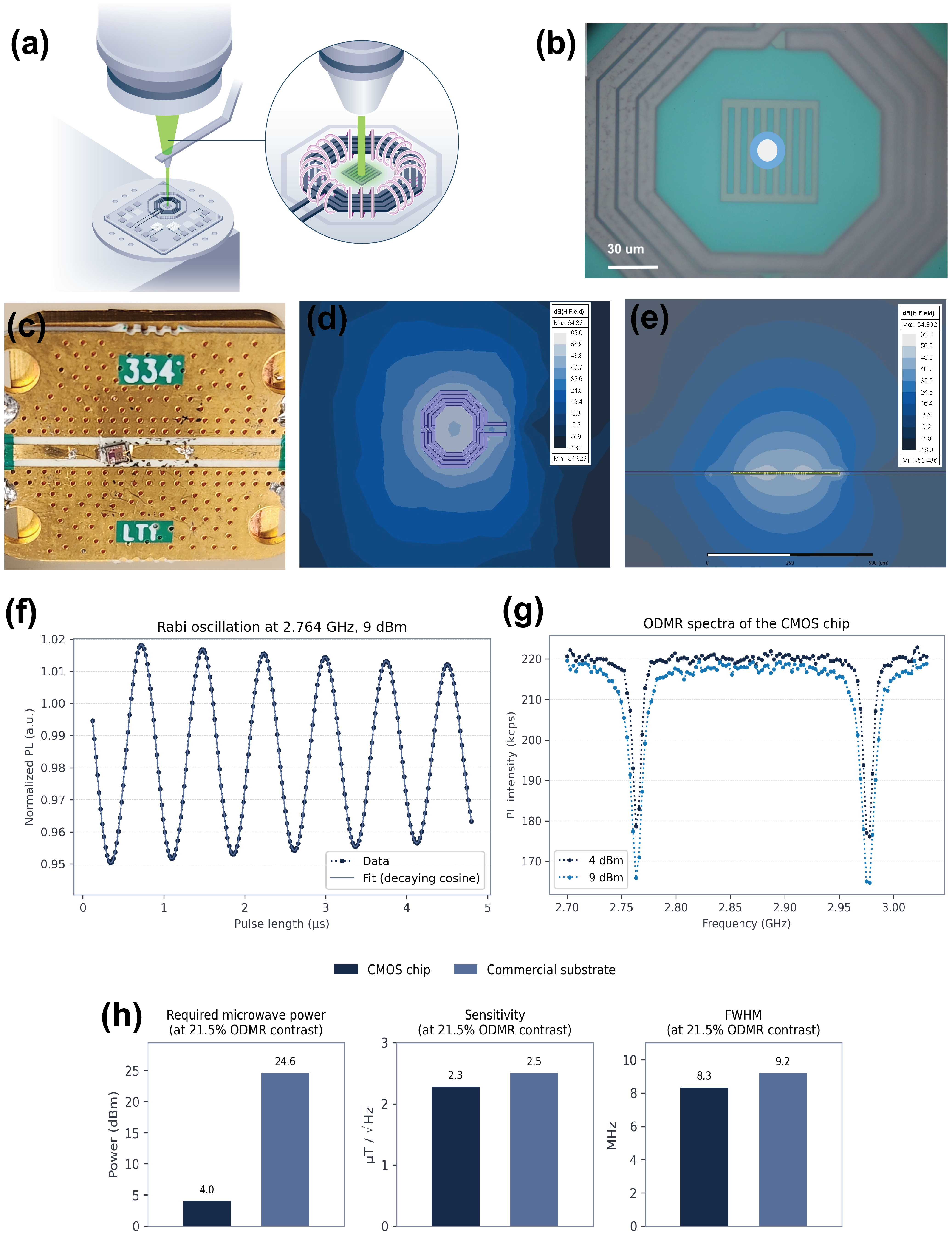}
    \caption{\textbf{Benchmarking the foundry CMOS platform against a commercial microwave antenna for low-power NV spin control.} (a) Schematic of the scanning NV measurement configuration, showing optical excitation and microwave delivery from the on-chip RF inductor. (b) Optical micrograph of the CMOS device highlighting the central inductor region used for spin control. (c) Experimental setup with the CMOS chip wire-bonded and interfaced for RF excitation. (d,e) Electromagnetic simulations of the RF magnetic field distribution generated by the on-chip inductor, showing spatial localization in the device plane and decay along the out-of-plane direction. (f) Rabi oscillations measured at 2.764~GHz, demonstrating coherent spin manipulation driven by the integrated RF field. (g) ODMR spectra acquired at 4~dBm and 9~dBm, showing high contrast ($>20\%$) at substantially reduced microwave power. (h) Performance comparison between the CMOS platform and a commercial antenna, including required microwave power, ODMR contrast, magnetic sensitivity, and linewidth (FWHM), highlighting efficient low-power operation with comparable sensing performance.}
    \label{fig:ODMR}
\end{figure*}

\begin{table*}[h]
\centering

\caption{Representative examples of CMOS-enabled quantum sensing architectures, reflecting a shared move toward foundry-compatible platforms. N/R = not reported in the referenced work.}
\label{tab:rf_platform_comparison}
\renewcommand{\arraystretch}{1.2}
\small
\begin{tabular}{p{4.5cm} p{2.8cm} p{2cm} p{3.5cm} p{1.7cm}}
\hline
\textbf{Platform} &
\textbf{RF architecture} &
\textbf{MW power (dBm)} &
\textbf{Key result} &
\textbf{Sensitivity $^{\dagger}$($\mu$T/$\sqrt{\mathrm{Hz}}$)} \\
\hline

Kim \textit{et al.} (CMOS-integrated NV sensor) \cite{kim2019cmos} &
Integrated MW delivery and photodetection &
N/R &
ODMR demonstrated &
32.1 \\

Ibrahim \textit{et al.} (High-scalability CMOS magnetometer) \cite{ibrahim2020high} &
Current-driven wire array &
N/R &
$<5\%$ MW field inhomogeneity &
0.245 \\

Fakkel \textit{et al.} (Cryo-CMOS NV controller) \cite{fakkel2024cryo} &
Integrated AC driver and bias coil &
N/R &
Rabi frequency up to 2.5 MHz &
N/A \\

Commercial antenna benchmark (this work) &
External antenna &
24.6--30.5 &
$\sim$20--25\% ODMR contrast &
$\sim$2.5 \\

\hline

\textbf{This work} &
\textbf{Integrated octagonal inductor} &
\textbf{4--9} &
\textbf{20--26\% ODMR contrast; Rabi demonstrated} &
\textbf{2.3--2.9} \\

\hline
\end{tabular}
\end{table*}

\vspace{0.3em}
\footnotesize
{
$^{\dagger}$ These works differ in target modality, RF architecture, and reporting conventions, so sensitivity values are not directly comparable.}

\normalsize

\subsection{In-situ electron-beam characterization}

Beyond optical and magnetic measurements, structural imaging is an important component of materials characterization. To evaluate compatibility with electron-beam-based workflows, we investigated device operation under direct SEM imaging conditions. The CMOS platform was mounted in an FEI Quattro environmental SEM (30~kV accelerating voltage, 630~pA beam current) and imaged while driving the integrated inductor \textbf{(Fig.~5a)}. Four bias voltages (0, 210, 730, and 2100~mV) were applied, corresponding to in-circuit currents spanning the microampere-to-milliampere regime. Because operation of the inductor generates both electric and magnetic fields, an important consideration is whether device actuation introduces imaging artifacts or degrades image quality. Representative SEM images acquired under different operating conditions are shown in\textbf{ Fig.~5b--d}, together with the corresponding fast Fourier transforms (FFTs). No visible image distortion, drift, or field-induced artifacts are observed. Likewise, the FFTs show no pronounced changes relative to the unbiased case, indicating that device operation does not introduce detectable periodic noise or imaging artifacts. The complete dataset, including the 2100~mV operating condition, is provided in Supplementary\textbf{ Fig.~S5.} To put this analysis on a more quantitative footing, we performed a knife-edge comparison of image quality across operating conditions. Sharp intensity transitions at the boundaries of the inductor features were used as probes of image resolution. If activation of the inductor introduced image degradation, a systematic reduction in edge sharpness would be expected with increasing bias. Edge-sharpness metrics extracted from the SEM images showed no statistically significant variation across the investigated operating conditions \textbf{(Fig.~5e--g)}. Details of the analysis procedure are provided in Supplementary Information: Knife-Edge Analysis section and Supplementary \textbf{Fig.~S5 - S11}. These results establish electron-microscopy compatibility as an additional capability of the CMOS platform, complementing the optical, electrical, thermal, and quantum sensing functionalities demonstrated above.

\begin{figure*}[htb!]
\centering
\includegraphics[width=0.9\textwidth]{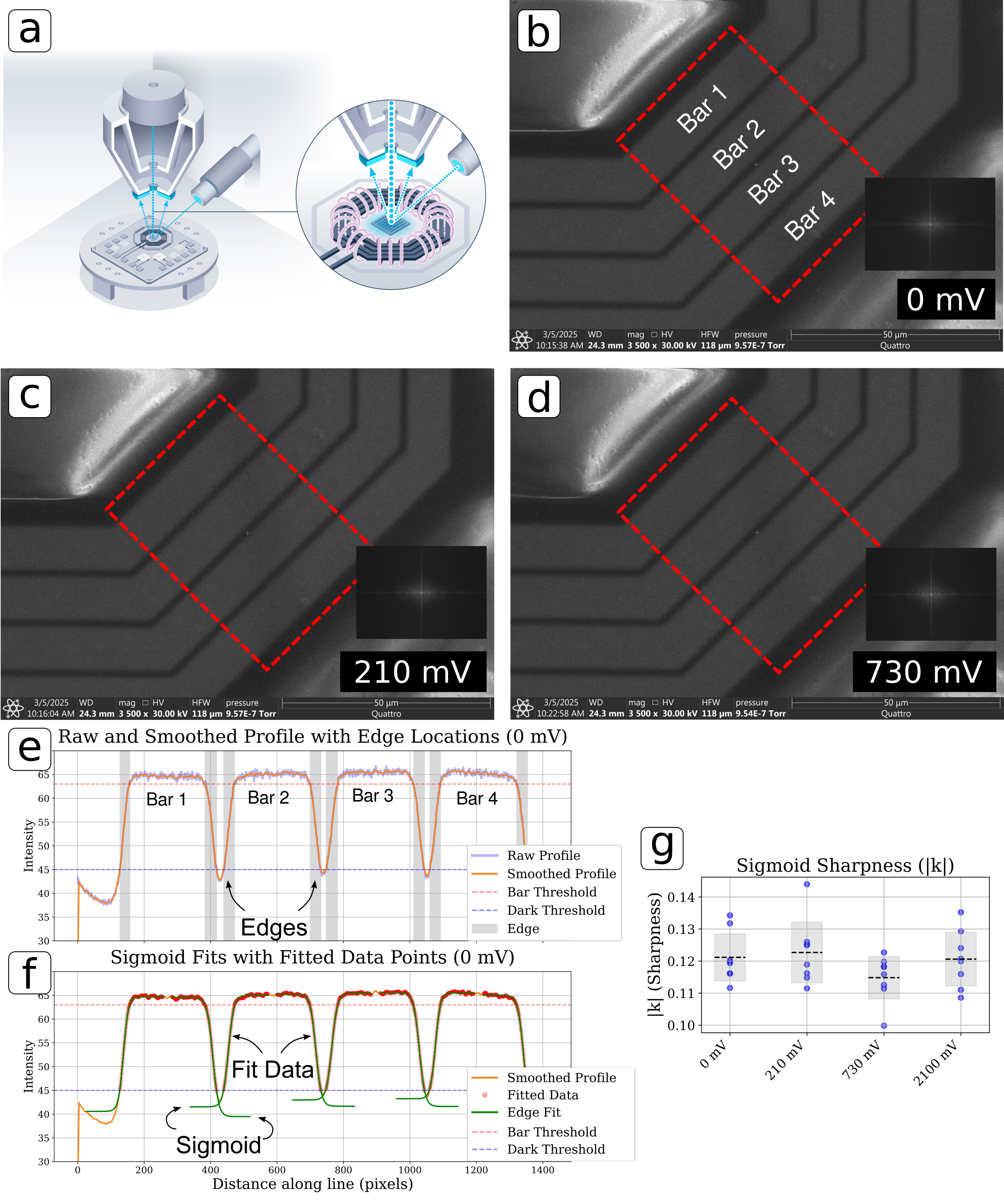}
\caption{\textbf{SEM compatibility of the CMOS platform during device operation.}
(a) Schematic of SEM imaging while the integrated inductor is electrically driven. (b--d) Representative SEM images acquired under different inductor bias conditions; corresponding FFTs are shown as insets. No visible image distortion or field-induced artifacts are observed during operation. (e) Overview of the knife-edge image-quality analysis used to quantify edge sharpness. (f) Representative sigmoid fits to the extracted edge profiles. (g) Extracted edge-sharpness metric for all operating conditions, showing no statistically significant variation with applied bias. Details of the analysis procedure are provided in the Supplementary Information (Knife-Edge Analysis section).
}

\label{fig4}
\end{figure*}

\section{Discussion and Outlook}\label{sec6}
While CMOS has become a mature technology for computation, communication, and sensing, its use as a standardized experimental substrate for quantum and materials characterization remains largely unexplored. The present work establishes the passive infrastructure layer of a progressively integrated characterization platform. Importantly, this first-generation implementation already demonstrates competitive performance relative to conventional characterization approaches. 

\begin{table}[t]
\centering
\caption{Functional partitioning of the foundry-fabricated CMOS characterization platform and corresponding experimental demonstrations.}
\label{tab:platform_summary}
\begin{tabular}{l l l p{3.0cm} p{3.8cm}}
\hline
Subsystem &
CMOS layer(s) &
Footprint &
Function &
Experimental validation \\
\hline
CMOS platform &
M1--M9 &
1 mm$^2$ &
Characterization platform &
Multimodal
\\

RF inductor &
M8--M9 &
212 $\mu$m outer span &
RF excitation &
RF characterization (RT, 4 K); ODMR; Rabi \\

IDE &
M7 &
$\sim$100 $\times$ 100 $\mu$m$^2$ &
Electrical interfacing &
I--V characterization; simulations \\

Resistive heater &
M5--M6 &
$\sim$100 $\times$ 100 $\mu$m$^2$ &
Thermal control &
TCR extraction; thermal conductance \\

\hline
\end{tabular}
\end{table}

The integrated RF architecture enables  cryogenic magnetic susceptibility measurements which demonstrate how the platform can be used as a reusable substrate for quantum materials studies. In contrast to conventional approaches that often require sample-specific device fabrication, excitation structures, or extensive processing steps, Fe$_3$GeTe$_2$ heterostructures were transferred directly onto the CMOS platform and characterized without additional lithography. This capability is particularly relevant for emerging low-dimensional and air-sensitive materials, where fabrication-induced variability and sample degradation can complicate measurements and limit reproducibility across experiments.

The platform further enables ODMR measurements with contrasts exceeding 20\% while operating at microwave powers of only 4–9 dBm, corresponding to a reduction of approximately 20–25 dB relative to conventional antenna-based implementations. Beyond reducing power consumption, the localization of field delivery within the CMOS stack enables a substantially smaller footprint than conventional microwave delivery hardware, highlighting the potential of foundry-integrated electromagnetic structures for compact sensing and characterization systems.

Recent advances in cryogenic CMOS technologies have demonstrated sophisticated active circuitry, including integrated microwave generation and delivery, pulse sequencing, qubit control, and scalable cryogenic readout architectures implemented in commercial 65\,nm technologies \cite{Tada2024,Das2022,Perez2023, Nagulu2023, Li2024}. These developments establish that active functionality can be realized within the same foundry platform employed in this work. Rather than reproducing these circuit capabilities, the present work addresses a complementary challenge: establishing the electromagnetic, thermal, electrical, optical, and heterogeneous integration infrastructure required for multimodal quantum materials characterization. This foundry-compatible characterization platform provides the foundation for future generations to directly incorporate existing cryo-CMOS circuit blocks, enabling fully integrated systems that combine active control, readout, and multimodal materials characterization within a unified architecture. The reproducibility afforded by foundry manufacturing also creates opportunities for automated experimentation, high-throughput characterization, and future machine-learning-assisted measurement and control.

\section{Methods}\label{sec11}

\subsection{Electrical Measurement}\label{subsec6.1}

\subsubsection{RF measurement}\label{subsubsec6.1.1}

The complex analysis of two-port S-parameters is performed using the vector network analyzer (Agilent N5225A) from 10\,MHz to 5\,GHz. The measurement setup included (i)~an empty test fixture and (ii)~the fixture loaded with the inductor device, both referenced to a characteristic impedance of 50\,$\Omega$ at 4\,K and room temperature.

To isolate the intrinsic device response, we employed a matrix-based de-embedding procedure. The measured S-parameters are converted to ABCD transfer matrices, and assuming a symmetric fixture, followed by deriving the half-fixture transfer matrix ($A_{\textit{half}}$) via the matrix square root of the empty fixture. Frequency-continuous branch selection was utilized to ensure sign consistency across the measured bandwidth. The de-embedded device matrix was computed as:
\begin{equation}
    A_{DUT} = A_{\textit{half}}^{-1} \, A_{Measured} \, A_{\textit{half}}^{-1}
\end{equation}
and subsequently converted back to the S-parameter domain. Numerical stability was maintained through diagonal regularization of ill-conditioned matrices using custom MATLAB scripts and the RF Toolbox. 

\subsubsection{Heater measurement}\label{subsubsec6.1.2}

The device-to-ambient thermal conductance $G_{\text{th}}$ (self-heating parameter) of the micro-heater is extracted electrically using four-terminal (Kelvin) I--V measurements, which were performed using a Keithley/Tektronix 2401 Source Meter to eliminate series resistance from pads and interconnects. The heater temperature coefficient of resistance (TCR) was first measured under conditions that suppress self-heating. Pulsed I--V measurements were performed while the ambient setpoint $T_a$ was varied from 25\,$^{\circ}$C to 125\,$^{\circ}$C, and the resistance was extracted from the plateau region of each pulse. The TCR was obtained from the slope of $R(T_a)$:
\begin{equation}
    \alpha = \frac{1}{R}\frac{dR}{dT}
\end{equation}
and reported at room temperature. Thermal coupling to the environment was then extracted by intentionally allowing self-heating under continuous (non-pulsed) bias. A linear voltage sweep was applied, and the corresponding current was recorded to compute $R = V/I$ and $P = IV$. The thermal conductance was calculated using $\Delta T = P / G_{\text{th}}$ combined with the calibrated $R(T)$, yielding:
\begin{equation}
    G_{\text{th}} = \left.\frac{dR}{dT}\right|_{T_a} \bigg/ \frac{dR}{dP}
    \quad \text{and} \quad
    R_{\text{th}} = \frac{1}{G_{\text{th}}}
\end{equation}
where $dR/dP$ was extracted from a linear fit of $R$ versus $P$ over mid--high-power points to avoid low-bias nonidealities. 

\subsubsection{IDE electrical measurement and simulation}\label{subsubsec8.1.3}

IDE electrical behaviors were quantified using complementary voltage-forcing and current-forcing I--V protocols on a Source Meter (Keithley/Tektronix 2401). In the voltage-forcing mode, the terminal voltage is swept from 0 to $V_{\max}$ (20\,V as the max limit of instrument) with a defined current compliance (typically $I_{\max}$), providing the leakage current $I(V)$ over the full available bias range. However, in current-forcing mode, the sourced current is swept from 0 to $I_{\max}$ (to the max limit of instrument,  1\,A  in this case) with a voltage compliance (set to the max limit of instrument), yielding the reciprocal transfer $V(I)$ and directly exposing high-impedance or leakage-limited behaviour without exceeding instrument limits. The two forcing modes were compared for consistency and used to identify the low-leakage regime (linear response and stable differential conductance).
Capacitance--voltage (C--V) measurements of the interdigitated
electrode were performed on a Keithley 4200-SCS equipped with the
4200-CVU module in a four-probe Kelvin configuration, with
H\textsubscript{CUR}/H\textsubscript{POT} and
L\textsubscript{CUR}/L\textsubscript{POT} pairs contacting the two
comb electrodes to eliminate lead and contact impedance. Open- and
short-circuit compensations were applied through the KTEI
environment prior to each run.

To connect electrical response to the spatial region probed above the IDE, finite-element electrostatic simulations (COMSOL) were performed using the as-designed electrode geometry and dielectric stack. The electric-field magnitude $|\mathbf{E}|$ was mapped above the IDE and reduced to a height-dependent profile $|\mathbf{E}(z)|$ normal to the surface, defining an effective sensing depth $z^{*}$ via:
\begin{equation}
    |\mathbf{E}(z^{*})| = \eta \, |\mathbf{E}(0)|
\end{equation}
The same model was used to compute the IDE capacitance under dielectric loading by replacing the medium above the electrodes with overlayers of controlled permittivity $\varepsilon_r$ and thickness $t$, yielding $C(\varepsilon_r, t)$ and the expected modulation $\Delta C$ for material placement on the IDE.

\subsection{Crystal growth}\label{subsec6.2}
 Fe$_3$GeTe$_2$ (F3GT) crystals were synthesized by chemical vapor transport (CVT). Elemental precursors were loaded into a quartz tube (10 mm in diameter and 200 mm in length) in the stoichiometric ratio Fe:Ge:Te = 3:1:2, together with I$_2$ as the transport agent. The tube was sealed under vacuum and placed in a two-zone furnace. The hot end containing the precursors was heated to 750$^\circ$C, while the cold end was maintained at 650$^\circ$C. The temperature gradient was kept for one week, after which the samples were naturally cooled to room temperature. Millimeter-sized crystals were found at the cold end of the tube. The single crystals were pre-characterized to assess their crystal quality and magnetic properties.

\subsection{Sample preparation}\label{subsec6.3}
Bulk single crystals were mechanically exfoliated onto polydimethylsiloxane (PDMS) films, doubled encapsulated between two hexagonal-BN (hBN) flakes of 20-30 nm thickness, and then transferred onto the CMOS chip using a polymer-based dry-transfer technique with stamps made of polycarbonate (PC) film covering a PDMS square [see Figure \textbf{ (Fig.~4b)}]. We glued the CMOS chip with epoxy on a standard PCB compatible with the cryostat employed for the measurements and the device was wire-bonded with Al wire.

\subsection{Sigmoid function}\label{subsec6.4}
The sigmoid function can be used to describe a step from one value to another value in a continuous manner. The sigmoid we used is given by

$$
    f(x) = \frac{L}{1+e^{-k(x-x_0)}} + A
$$

\noindent where $L$ controls the scale, $A$ controls the $y$-offset, $x_0$ controls the position, $|k|$ controls the sharpness of the transition, and the sign of $k$ controls whether it is a rising edge (low to high, $k>0$) or a falling edge (high to low, $k<0$) with increasing $x$. Fitting sigmoid functions of this form allows us to obtain a sharpness metric, $k$, for each edge and compare the results from each image. Examples of sigmoid functions fit to the edges are shown in \textbf{Fig.~5f} in the lower panel. The portion of the data used for fitting is shown in red; due to the cyclic nature of the profile, it is inappropriate to use the full dataset for fitting. We confine our fit to the edges, while including a portion of the flat area on top of the inductor bars to ensure the best fit. A summary and comparison of the extracted $|k|$ values, along with the mean (dotted lines) and standard deviation (gray boxes), is given in \textbf{Fig.~5g}. From this data we conclude that there is no observable degradation in image quality on activation of the inductor. 

\subsection{Optical measurements}\label{subsec6.4}
Magnetic circular dichroism (MCD) were recorded in back scattering geometry and measured using a standard photoelastic modulator (PEM, Hind's Instruments) technique by modulating the incident beam polarization with $\lambda/4$ retardance at a frequency of 50 kHz. The differential signal at the PEM frequency was measured by a lock-in amplifier (Stanford Research Systems, SR865A). Reflectance measurements were simultaneously recorded using a separate lock-in amplifier (SR830) and a mechanical chopper ($f = 781$ Hz). This reflectance signal is also used as a normalization for the MCD data. We used a supercontinuum laser source (NKT Photonics) and a home-made custom monochromator that provides light in the range 1.3 eV to 2.8 eV. A 50x objective lens (Olympus) was used to focus the beam onto the sample (with spot size $\sim$2-4 $\mu$m, dependent on wavelength). The Kerr rotation was measured modulating the incoming linearly polarized light with a mechanical chopper ($f = 781$ Hz). The detection scheme consisted in a half-wave plate (HWP), a Wollaston prism (WP) and a balanced photodetector (BPD) (ThorLabs PDB210A, bandwidth DC - 1 MHz). The HWP was used to calibrate the BPD on a spot nearby the flake of interest. The differential signal at the chopper frequency was measured by a lock-in amplifier (Stanford Research Systems, SR865A). For the susceptibility measurements the current was injected into the micro-inductor from the output of the same lock-in amplifier, while the other end was connected to a 50 $\Omega$ resistor. Typical incident powers were between $1$-$100$ $\mu$W. Temperature and magnetic field dependent optical measurements were performed in a Quantum Design Opticool cryostat with a base temperature of $\sim 1.75$ K and in magnetic fields up to 3 T.

\section*{Acknowledgements}

S.K.Y. and J.A.  thank the KAUST Core Labs staff: Camelia Florica and Qingxiao Wang for SEM and cross-sectional SEM imaging, respectively, and Syed Kazmi and Menouer Saidani for assistance with etching the chips. J.A. and L.N. thank Dr. Bevin Huang for the discussion on the magnetic susceptibility measurements. J.A. acknowledges the support from KACST-MIT Ibn Khaldun Fellowship for Saudi Arabian Women at MIT, from Ibn Rushd Postdoctoral Award from King Abdullah University of Science and Technology (KAUST), and from the Army Research Office MURI (Ab-Initio Solid-State Quantum Materials) Grant No. W911NF-18-1-043.

\section*{Data availability}

All data supporting the findings of this study are available from the corresponding author (J.A.) upon request.

\printbibliography

\end{document}


\renewcommand{\thefigure}{S\arabic{figure}}
\setcounter{figure}{0}

\maketitle

\newpage

\section{Transmission coefficient data smoothing}

S11 parameters were measured from 0.1 to 5.0 GHz at both 300 K and 4 K using a vector network analyzer. The raw data were smoothed using Savitzky–Golay digital filters with  window length = 51 points (50 MHz), polynomial order 3 for the 4 K data (1 MHz step, 4901 points); window length = 11 points (196 MHz), polynomial order 3 for the 300 K data (19.6 MHz step, 251 points). Window sizes were chosen proportional to measurement density to achieve comparable effective smoothing bandwidths. Similarly, the S21 raw data were smoothed using Savitzky–Golay digital filters with  window length = 51 points (50 MHz) for both 300 K data and 4 K data.

\begin{figure*}[h]
\centering
\includegraphics[scale=0.63]{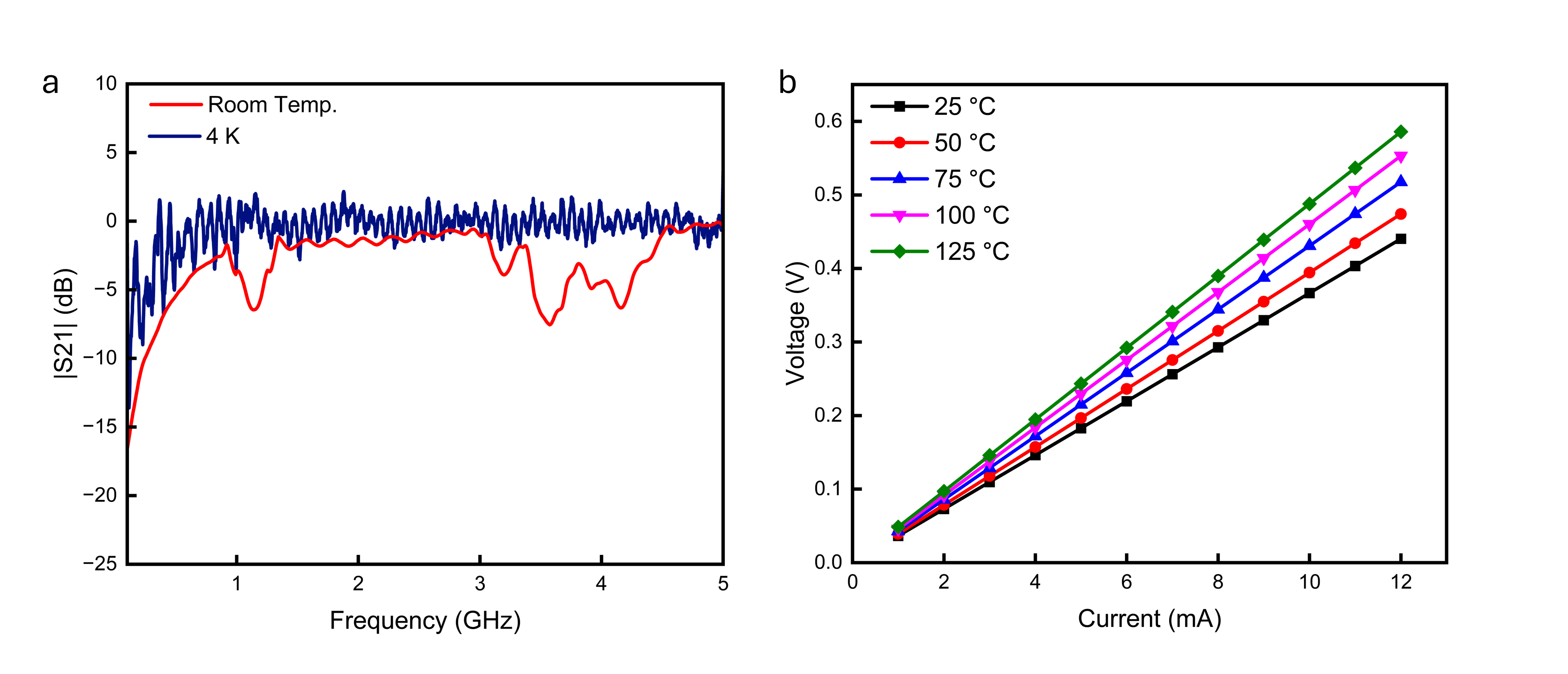}
\caption{\textbf{(a)} Broadband microwave transmission of the on-chip four-turn octagonal loop inductor at room and cryogenic temperatures. De-embedded magnitude of the transmission coefficient |S21| from 10 MHz to 5 GHz at 300 K (red) and 4 K (blue), referenced to \SI{50}{\ohm}. |S21| remains close to -1 dB at 4 K and -3 dB at room temperature across the operating bandwidth, indicating higher power transmission and lower dissipative loss. The fine ripple at 4 K originates from standing waves in the cryogenic line. The close agreement between traces confirms that the microwave performance is preserved upon cooldown. \textbf{(b)} DC current--voltage characteristics of the heater measured at ambient temperatures of 25, 50, 75, 100 and \SI{125}{\celsius}. The strictly linear $V$--$I$ response confirms ohmic conduction over the full \SIrange{1}{12}{\milli\ampere} bias range, with no sign of self-heating (as a pulsed current excitation was used to suppress Joule self-heating and isolate the intrinsic temperature dependence)  or nonlinearity. The slope, corresponding to the series (winding) resistance, increases monotonically with temperature reflecting the positive temperature coefficient of resistance ($3.36\times10^{-3}\,\mathrm{K^{-1}}$) characteristic of the metal traces. The uniform, well-separated isotherms demonstrate stable and predictable resistive behaviour across the tested temperature span.}
\label{|S21|}
\end{figure*}

\section{IDE electrical measurement}

\begin{figure}[H]
\centering
\includegraphics[scale=0.35]{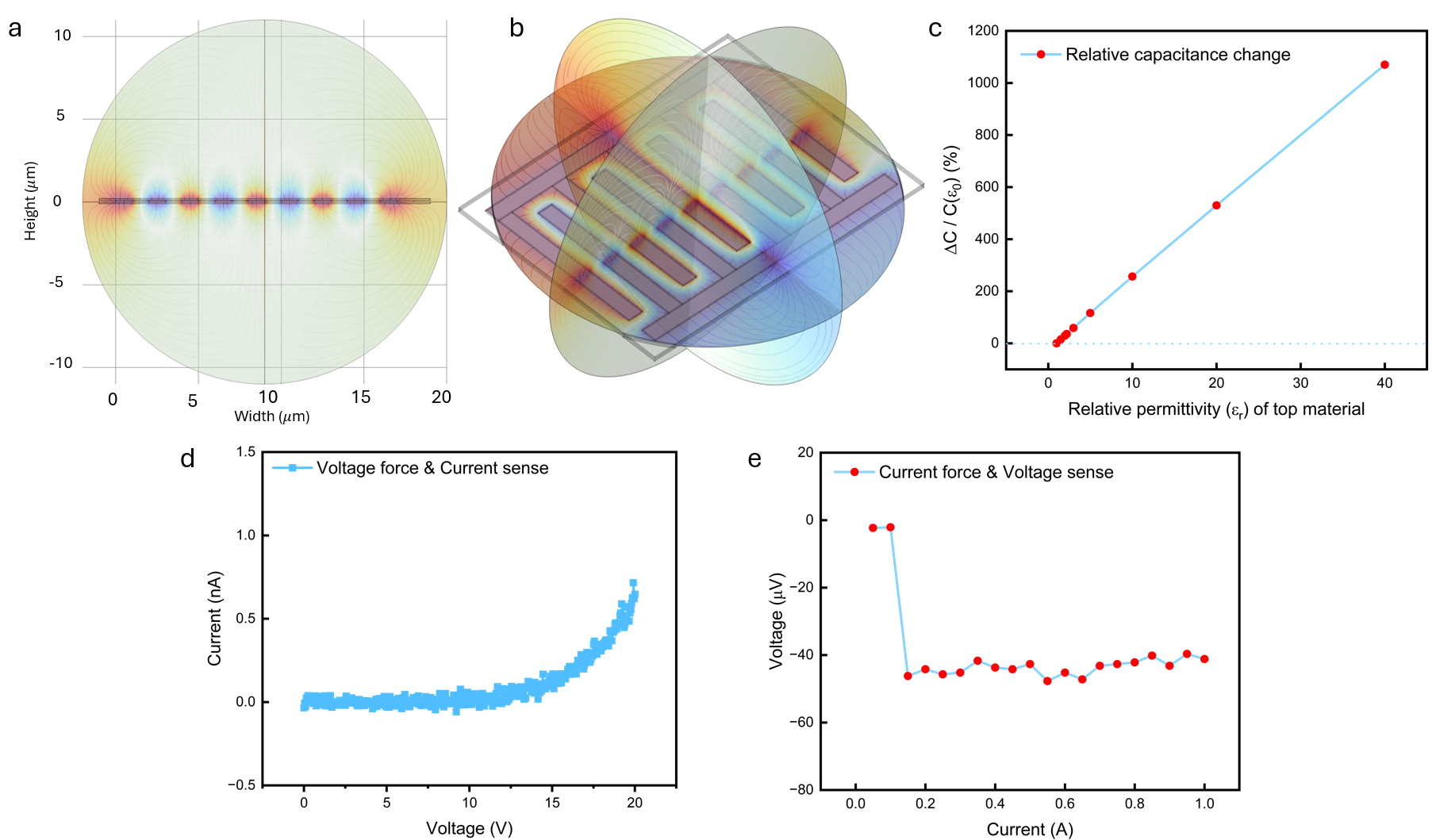}
\caption {Finite-element electrostatic modeling and experimental electrical characterization of the IDE. (a) Simulated cross-sectional electric field distribution above the electrode surface; the field magnitude |E(z)| decays with distance from the surface, defining an effective sensing height set by the electrode geometry and confining the field to the near-surface region. (b) Three-dimensional rendering of the simulated field distribution over the interdigitated finger array, illustrating localized probing of dielectric perturbations in the immediate vicinity of the IDE. (c) Simulated relative capacitance change, $\Delta C_{\mathrm{sim}}/C_{\mathrm{sim}}(\varepsilon_{r}=1)$, versus relative permittivity $\varepsilon_{r}$ of the top overlayer material, normalized to the baseline capacitance in air ($\varepsilon_{r}=1$); the response increases linearly with permittivity, reaching ${\sim}1070\%$ at $\varepsilon_{r}=40$. 
(d) Measured voltage-forcing/current-sensing characteristic, I(V), with voltage swept from 0 to 20 V, showing leakage current below 50 pA up to ~12 V before the onset of conduction. (e) Measured current-forcing/voltage-sensing characteristic, V(I), with sourced current swept from 0 to 1 A, exposing the high-impedance regime. Together, panels (d)–(e) define the low-leakage bias window for reliable electrical interfacing and dielectric sensing.}

\label{IDE}
\end{figure}

\section{Kerr Rotation}
\begin{figure}[H]
\centering
\includegraphics[scale=1.8]{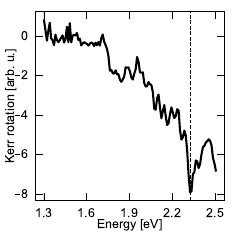}
\caption{Kerr rotation signal measured as a function of the incident light energy, in the range 1.3 to 2.5 eV. The energy selected for the measurements presented in the main text is 2.33 eV (533 nm), which represents the maximum in the Kerr rotation spectrum.}
\label{kerr}
\end{figure}

\section{Commercial antenna benchmarking}

\begin{figure}[H]
\centering
\includegraphics[scale=0.3]{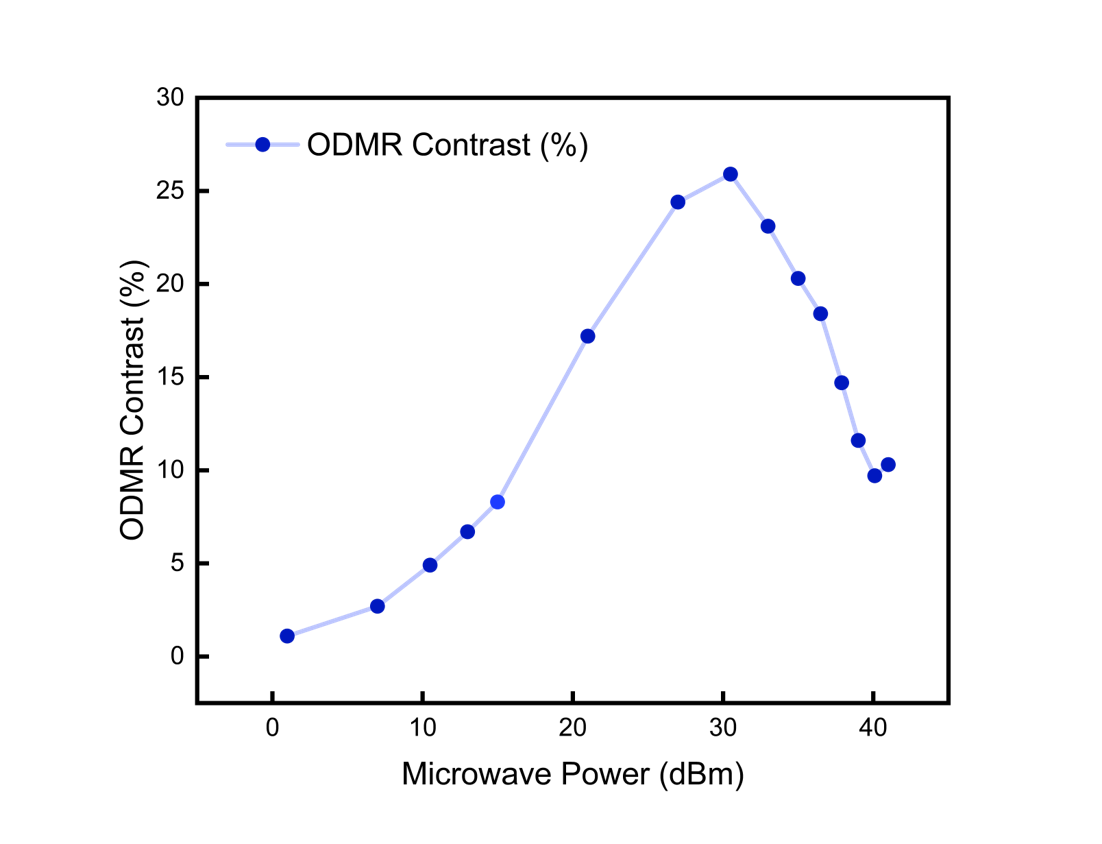}
\caption{Systematic microwave-power sweep of the commercial microwave antenna used for benchmarking against the CMOS platform. The complete power sweep establishes the operating characteristics of the commercial antenna and provides the basis for the representative comparison presented in Fig.~4h of the main text.}
\label{fig:commercial_antenna}
\end{figure}

\section{Knife-Edge Analysis}
In the main text, we gave a condensed presentation of a knife-edge analysis of SEM images to assess relative image quality and determine if there is any image degradation upon activation of our inductor coil. Here, we present a more detailed version of this analysis.

\begin{figure*}[h]
\centering
\includegraphics[scale=0.1]{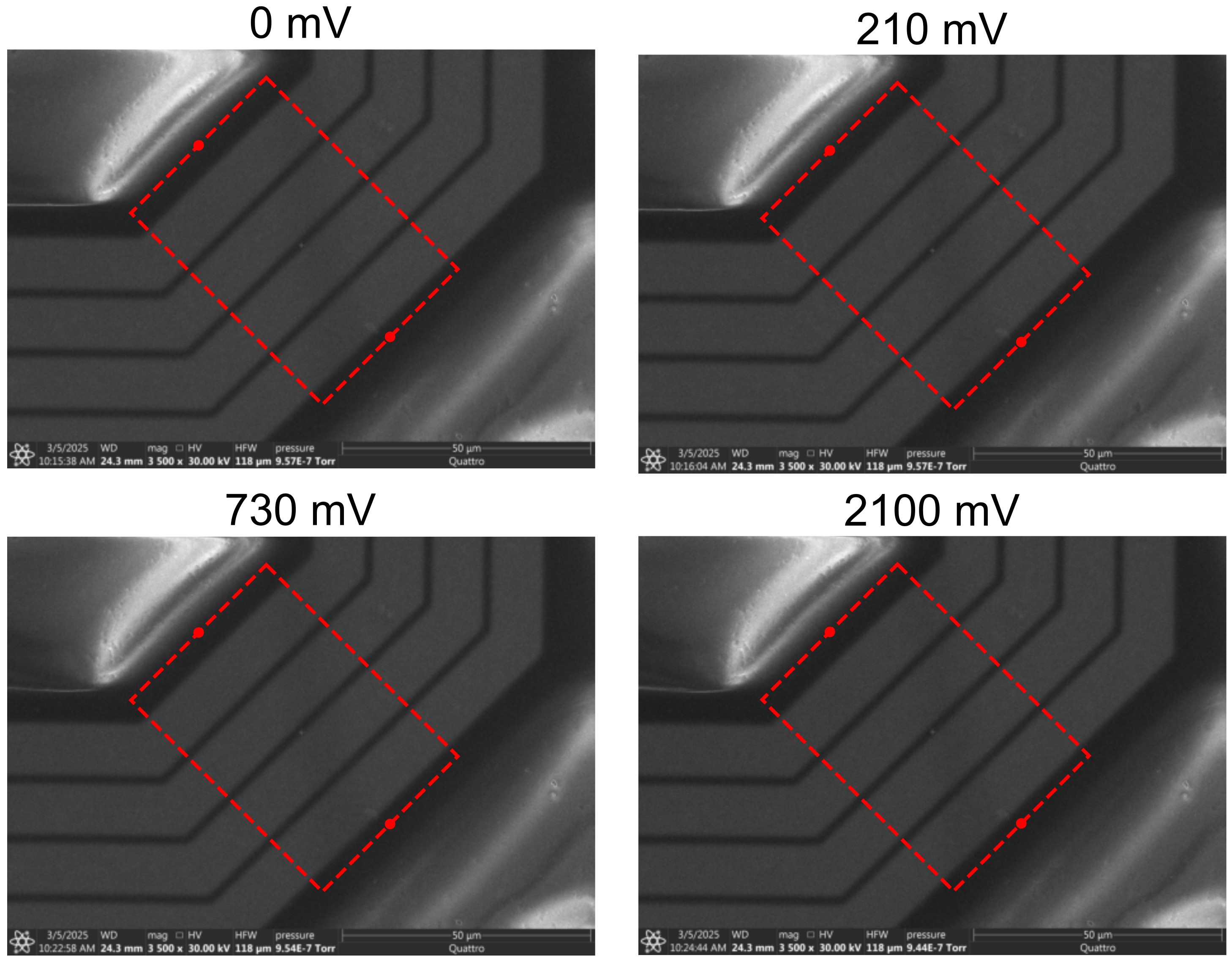}
\caption{SEM images of inductor coil at various excitation strengths. Dotted, red boxes represent the location of the extracted intensity profiles.}
\label{SEM_images}
\end{figure*}

Four images were acquired in an FEI Quattro environmental SEM using an accelerating voltage of 30 kV and a beam current of 630 pA. The inductor coil was activated with voltages of 0, 210, 730, and 2100 mV. The associated images are shown in Fig. \ref{SEM_images}. Based on visual inspection, there is little difference between them with regard to both image content (structure) and imaging quality (feature sharpness). Here, we perform a knife-edge analysis to quantify this qualitative conclusion based on visual inspection.

A knife-edge analysis of image quality leverages the existence of a sharp discontinuity within the field of view to assess the ability of the imaging system to reproduce this discontinuity. For example, taking a picture of a literal knife-edge that is back-lit will allow an assessment of the resolution of the optical imaging system used to produce the image.

Here, we are not interested in quantifying the overall image quality, but are looking for any relative changes in quality between the images that can be attributed to the activation of the inductor coil. In other words, we would like to know if there is a trend toward decreasing image quality with an increase in the voltage applied to the inductor.

The metal traces of the inductor coil, Fig. \ref{SEM_images}, that have been imaged provide us with a series of sharp transitions from bright to dark intensity. These transitions allow us to perform a knife-edge analysis and obtain a quantitative comparison of image quality.

\begin{figure*}[]
\centering
\includegraphics[scale=0.13]{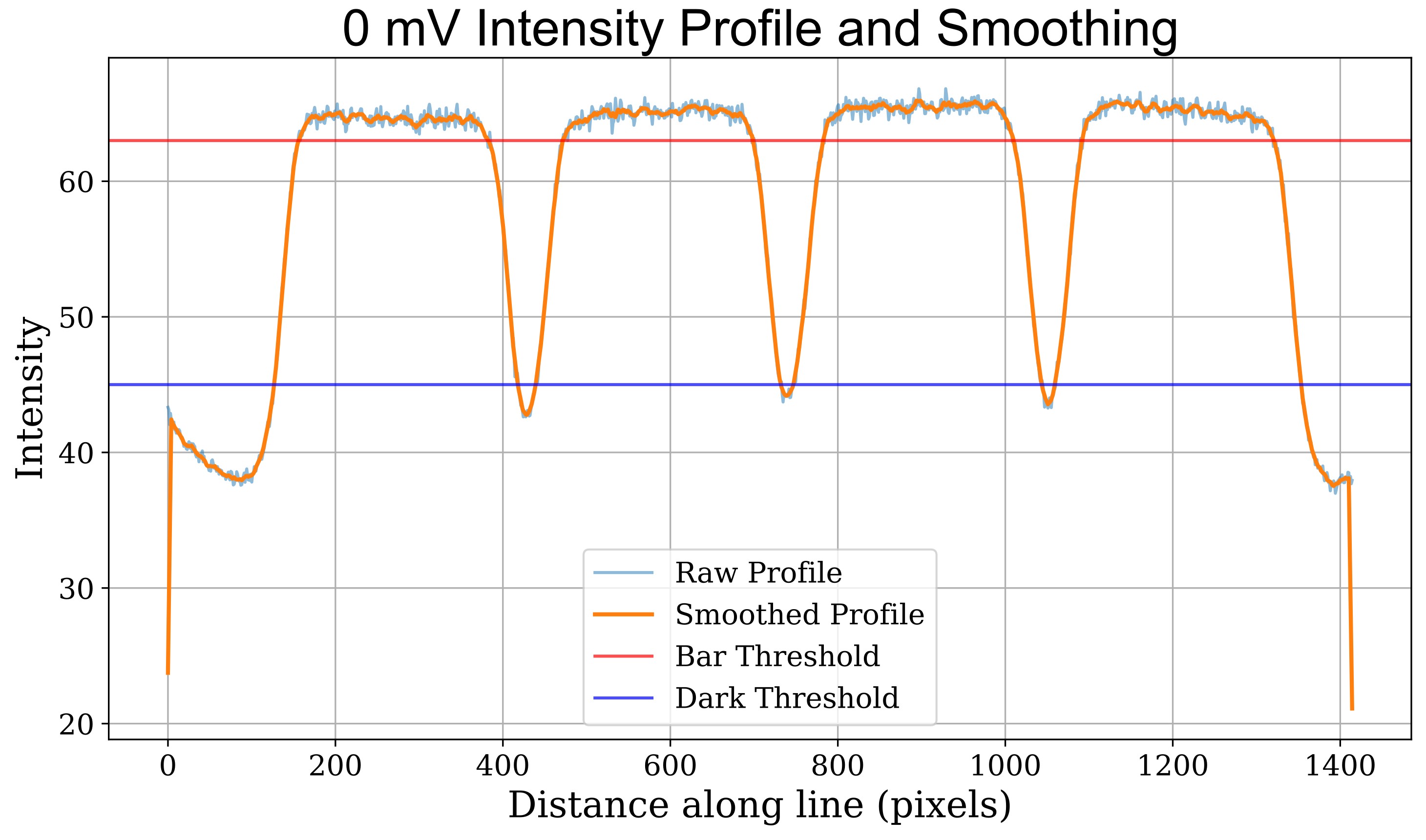}
\caption{Example intensity profile from the 0 mV image. The raw (averaged) signal was smoothed with an 8-pixel rolling average. High and low intensity thresholds were manually selected so that the approximate edge locations could be determined based on threshold crossing.}
\label{profile}
\end{figure*}

\begin{figure*}[]
\centering
\includegraphics[scale=0.13]{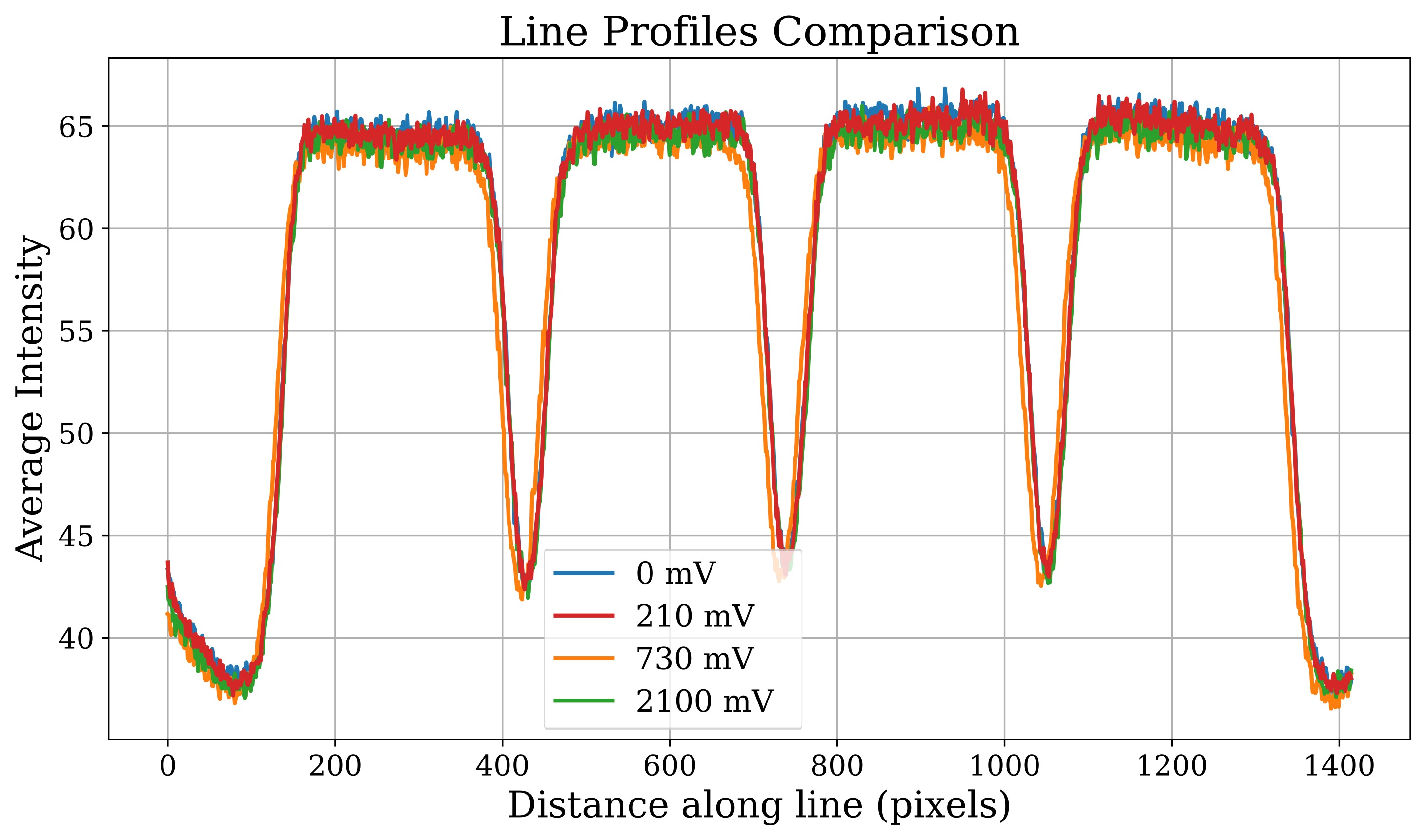}
\caption{Comparison of all the intensity profiles (before smoothing).}
\label{all profiles}
\end{figure*}

\begin{figure*}[]
\centering
\includegraphics[scale=0.07]{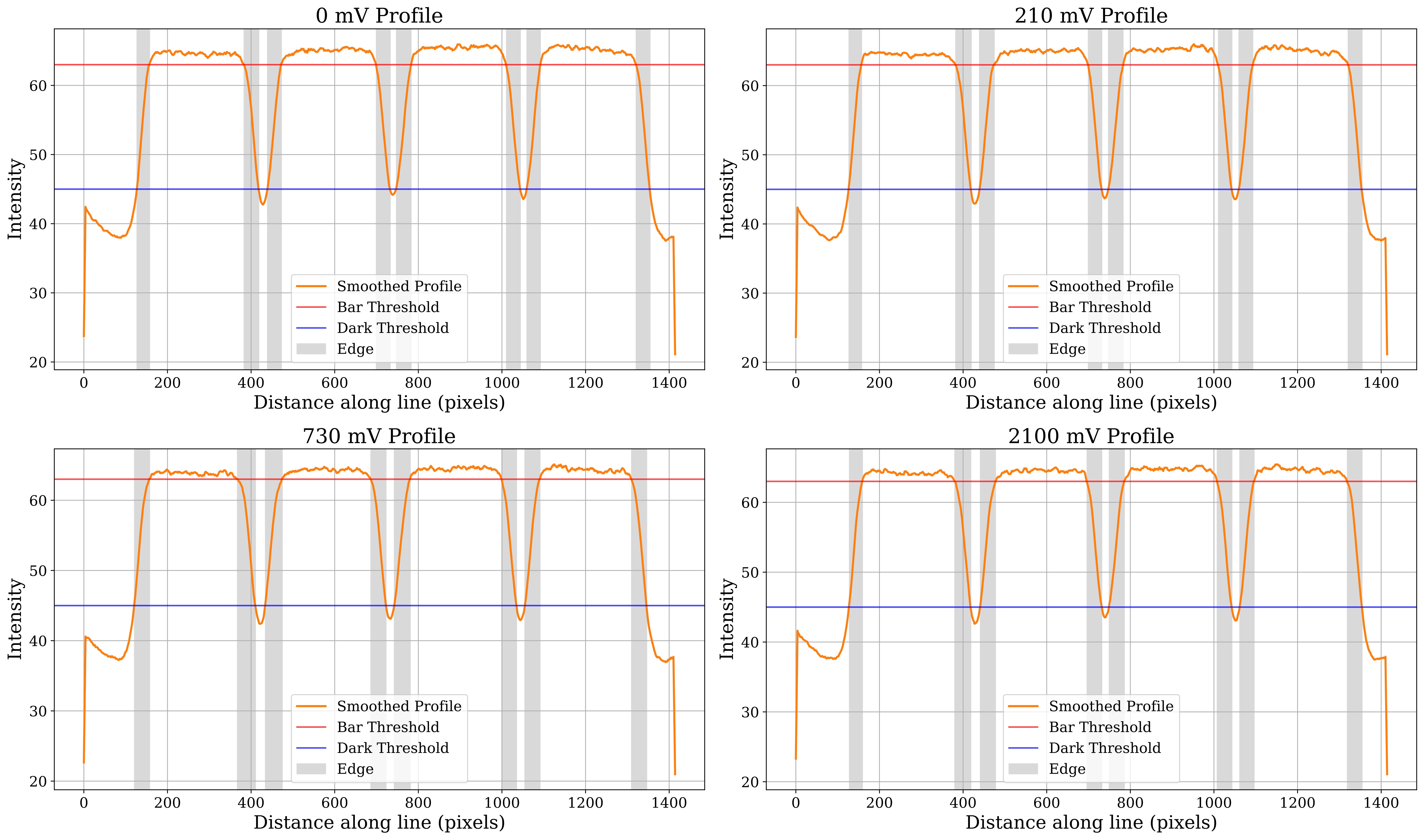}
\caption{Edges detected using manual thresholding. Edges are defined to be those regions of the intensity profile that are above the dark threshold (blue horizontal line) and below the bar threshold (red horizontal line). Regions classified as edges are indicated by the grayed areas.}
\label{threshold}
\end{figure*}

The first step toward performing the knife-edge analysis is to extract the intensity profile across the inductor coil. The regions chosen are indicated by the boxed regions in Fig. \ref{SEM_images}.  The width of the boxed region is 1000 pixels, giving 1000 intensity profiles, which are averaged together to obtain a low-noise, average intensity profile. An 8-pixel rolling average window was additionally applied to further reduce noise. The average intensity profile and the smoothed profile are shown in Fig. \ref{profile}. We do not want to aggressively smooth the intensity profile for fear of artificially reducing the final image quality measurement\textemdash smoothing our knife edge\textemdash but a modest 8-pixel smoothing window nicely reduces small noise spikes without significantly altering the edge profiles, as can be seen in Fig. \ref{profile}. 
Next, we want to reliably identify the location and approximate width of the observed edges. To do this we manually selected two thresholds, shown in Fig. \ref{profile}\textemdash one for the higher, 'bar' intensity of the inductor coil bars, and one for the lower intensity surrounding the bars. These are labeled 'Bar Threshold' and 'Dark Threshold' and represented as horizontal lines in Fig. \ref{profile}. These thresholds were manually selected so that they would work across all the intensity profiles. A comparison of the average profiles for all four images prior to smoothing is shown in Fig. \ref{all profiles}.
Using the intensity thresholds defined in the previous step, we defined the approximate position, width, and height of each observed edge. Fig. \ref{threshold} shows the results of this procedure, where the detected edges for all four profiles are shown using gray boxed overlays.
With the approximate positions, widths, and heights of the edges we can now fit an analytic function to the edge profiles in order to extract a quantitative metric of the edge sharpness for each edge. For this purpose we chose the sigmoid function, given by the expression
$$
f(x) = \frac{L}{1+e^{-k(x-x_0)}} + A
$$
\noindent where $L$ controls the scaling, $A$ controls the $y$-offset, $x_0$ controls the $x$ position, $|k|$ controls the sharpness of the transition, and the sign of $k$ controls whether it is a rising edge or a falling edge (i.e. low to high or high to low).\\

\begin{figure*}[]
\centering
\includegraphics[scale=0.13]{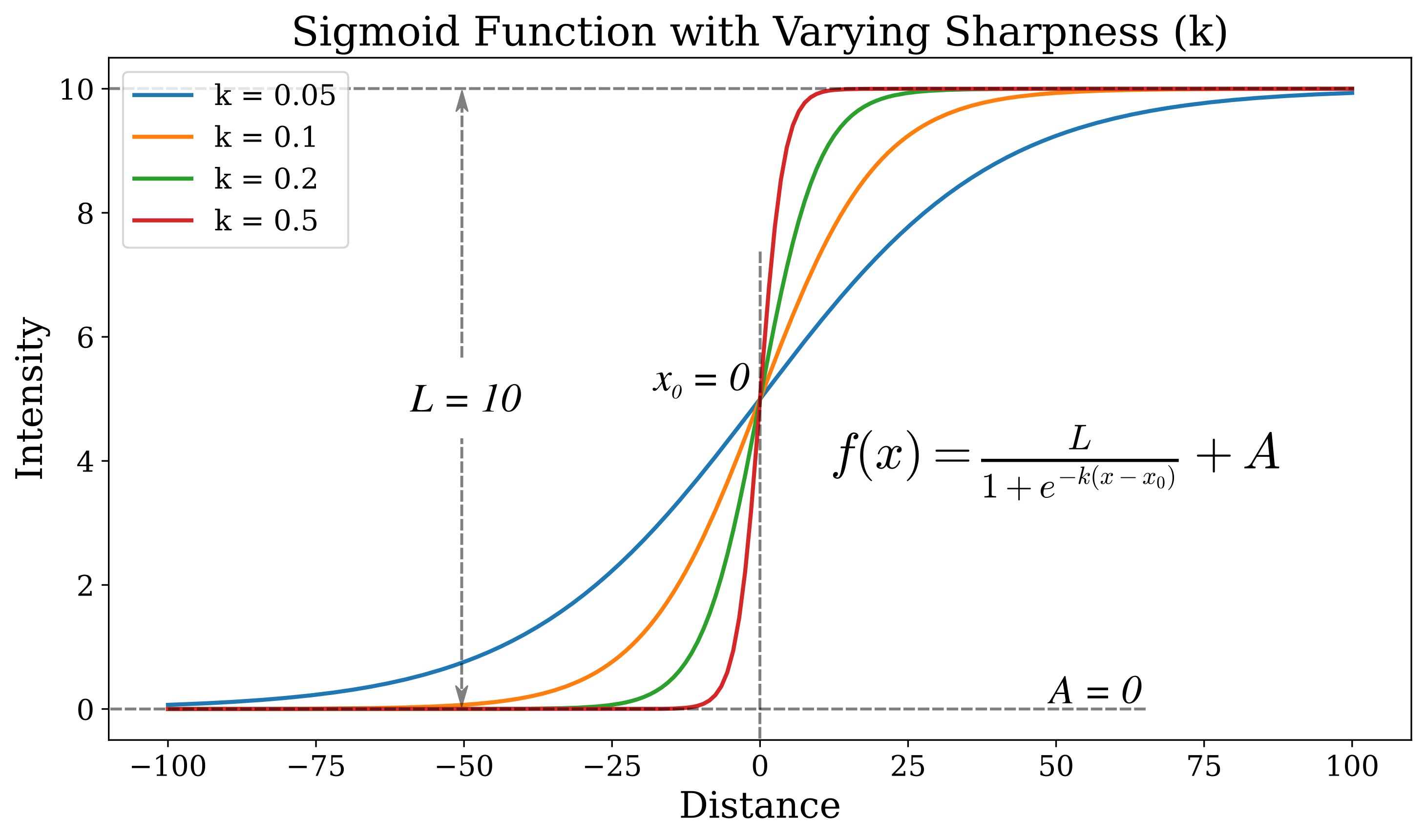}
\caption{Illustration of the impact of the $k$ parameter on the sharpness of the sigmoid function.}
\label{sigmoid}
\end{figure*}

 Fig. \ref{sigmoid} illustrates the behavior of a sigmoid of this form for various values of $k$. The other parameters were set as follows: $L=10$; $A=0$; $x_0=0$.

As we can see, increasing the value of $k$ increases the sharpness of the edge. Fitting a function of this form to our edges will allow us to quantitatively compare the edge sharpness for different images and edges.

A further consideration arises when trying to decide on an appropriate fitting window. The sigmoid function is asymmetric and non-periodic, whereas our intensity profiles are (roughly) both periodic and symmetric. So, we must decide on an appropriate fitting window that will allow the sigmoid to accurately approximate each edge without including datapoints from neighboring areas that would appreciably disturb the fit. We note that the dark regions on either side of the inductor coils (at $x=100$ and $x=1400$) are darker than the dark regions between the inductor coils. This indicates that the metal bars of the inductor are close enough together that the dark intensity between the bars has not reached its minimum value before the onset of the edge of the next bar. For this reason, we do not have data to reliably fit the lower (dark) portion of the sigmoid. However, each bar does provide us with a relatively uniformly flat, bright area along the top. To ensure that at least one of the asymptotes for the sigmoid is represented in the fit, we chose to include an additional 100 pixels beyond the rough estimate for the edge location in the bright regions\textemdash beyond the bar threshold. In the dark regions, we truncated the fitting window at the lower threshold to prevent the sigmoid from fitting to data from the adjacent edge.

\begin{figure*}[]
\centering
\includegraphics[scale=0.08]{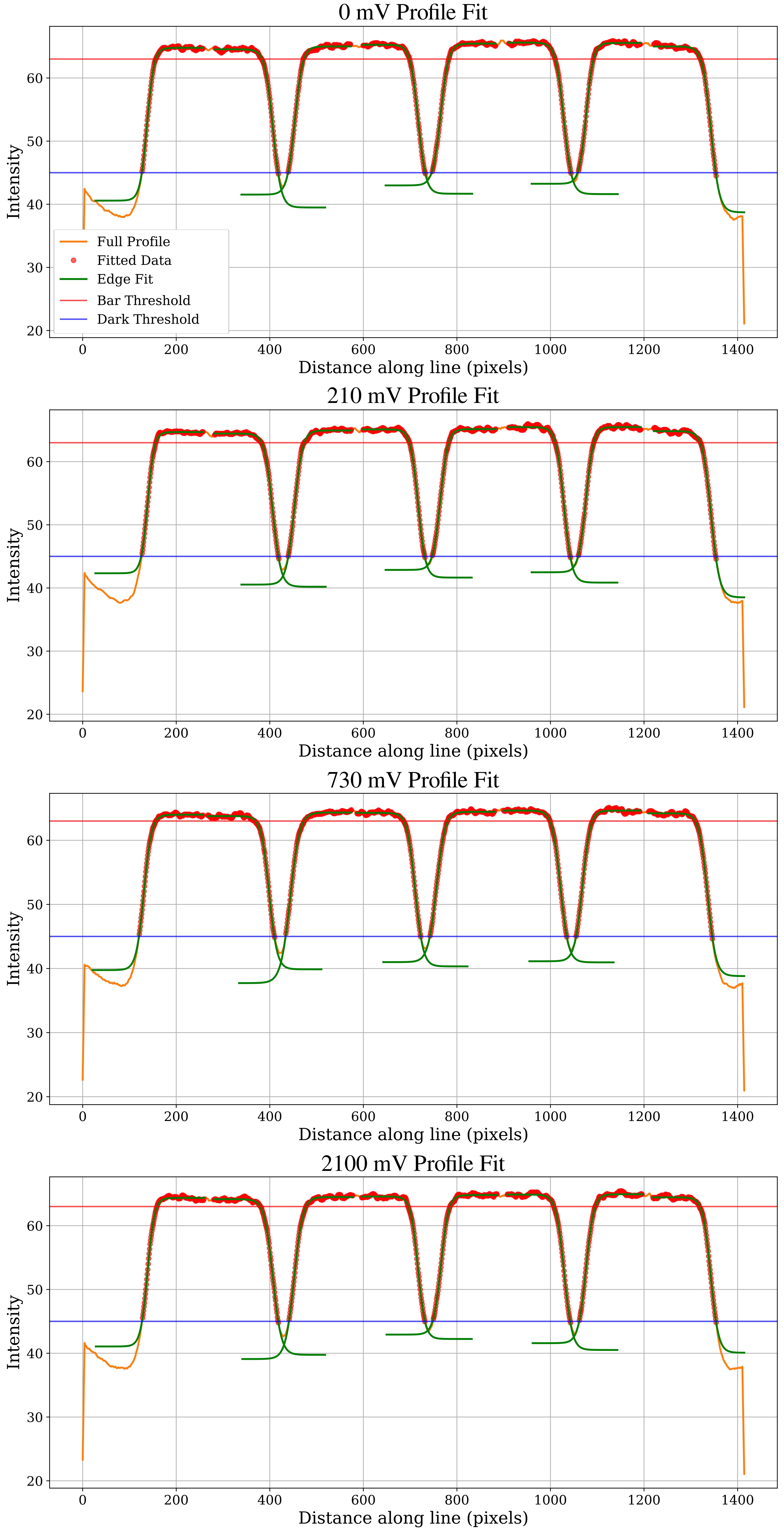}
\caption{Fitting results. The smoothed intensity profile is shown in orange. The bar and dark thresholds are indicated by the horizontal red and blue lines, respectively. The data points used for fitting are shown as red dots. The sigmoid fits are shown as green curves.}
\label{fit}
\end{figure*}

\begin{figure*}[t]
\centering
\includegraphics[scale=0.19]{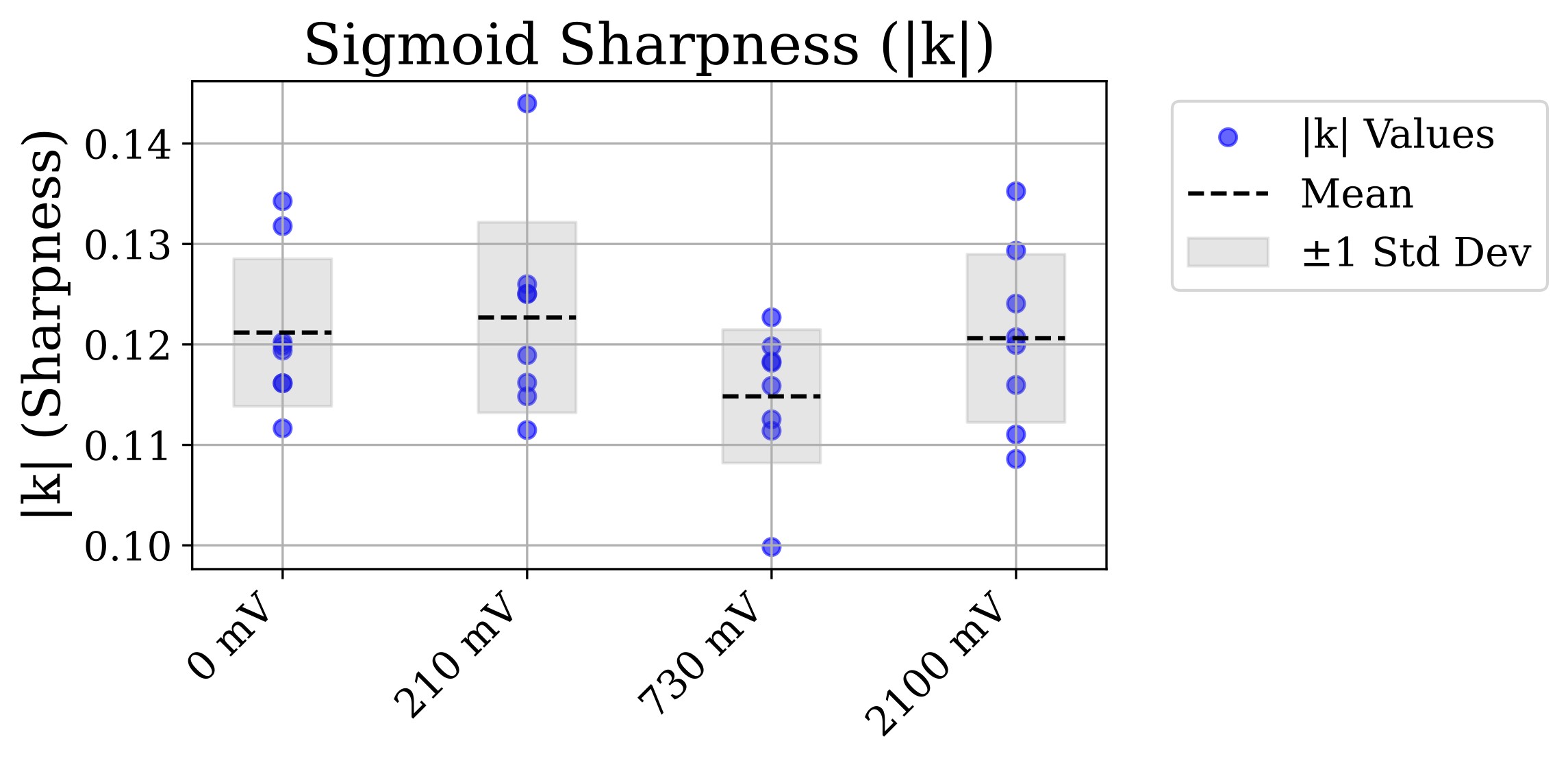}
\caption{Comparison of the $|k|$ values for all fits. Blue dots represent all $|k|$ values measured. Dotted lines indicate the mean. Gray boxes indicate $\pm$ one standard deviation from the mean.}
\label{k_values}
\end{figure*}

The results of these fits are shown in Fig. \ref{fit}. The full, smoothed profiles are shown in orange, with the data points used for fitting shown in red. The optimized sigmoid curves are shown in green. The sigmoid curves are plotted beyond their fitting range in the dark regions so that we can make a qualitative judgment as to the quality of the fit. Because there is no dark asymptote available for fitting, it is possible that some fits could extend unrealistically low. By plotting the sigmoid curves beyond the dark fitting region, we can obtain predictions for the estimated dark asymptote and compare the agreement between the different fits. By visual inspection we can see that, while there is some variability in the dark asymptote prediction between the fits, they are all in general agreement and all suggest a believable dark level that is roughly in agreement with the dark level at the edges of the inductor and slightly below the dark level observed between the coils. In other words, the predicted dark levels are as expected with none of the images displaying any anomalous fits that would indicate cause for concern in our edge sharpness comparison. 
Now that we are confident that our sigmoid fitting procedure has produced a sensible approximation for each of our edges, we can generate a scatter plot of all of the $|k|$ values obtained. The plot shown in Fig. \ref{k_values} gives this comparison. The dotted lines indicate the mean edge sharpness for each image and the gray boxes indicate $\pm$ one standard deviation from the mean. We conclude that there is no trend and that our quantitative measure of image quality detects no degradation in image quality with the activation of the inductor.

Python code that reproduces this analysis is available on GitHub at \url{https://github.com/ondrejdyck/Knife-Edge-Analysis}

